# Mechanistic Transition from Phonon Propagation to Thermal Hopping in Two-Dimensional Solids

Yanlei Wang[1], Zhigong Song[1], Zhiping Xu[1,2*]

[1]Applied Mechanics Laboratory, Department of Engineering Mechanics and Center for Nano and Micro Mechanics, Tsinghua University, Beijing 100084, China

[2]State Key Laboratory of Mechanics and Control of Mechanical Structures, Nanjing University of Aeronautics and Astronautics, Nanjing 210016, China.

[*]Corresponding author, Email: xuzp@tsinghua.edu.cn

**Thermal transport in solids changes its nature from phonon propagation that suffers from perturbative scattering to thermally activated hops between localized vibrational modes as the level of disorder increases. Models have been proposed to understand these two distinct extremes that predict opposite temperature dependence of the thermal conductivity, but not for the transition or the intermediate regime. Here we explore thermal transport in two-dimensional crystalline and amorphous silica with varying levels of disorder, $\alpha$, by performing atomistic simulations as well as analysis based on the kinetic and Allen-Feldman theories. We demonstrate the crossover between the crystalline and amorphous regimes at $\alpha \sim 0.3$, which can be identified by a turnover of the temperature dependence in thermal conductivity, and explained by the dominance of thermal hopping processes. The determination of this critical disorder level is also validated by the analysis of the participation ratio of localized vibrational modes, and the spatial localization of heat flux. These factors can serve as key indicators in characterizing the transition in heat transport mechanisms.**

Unraveling the correlation between thermal transport in solids and their microstructures has been a continuing effort undertaken from both fundamental and application standpoints. Models have been proposed to describe thermal conduction at the microscopic level, which can be classified into two major categories. Perfect crystals feature translational lattice symmetries and thus the language of phonon applies.[1] Phonon dispersion and interaction parameters extracted from the equilibrium crystal structure can be used to derive kinetic models. Predictive calculations can then be done based on the specific heat, phonon group velocities, and rates of scattering processes following the Boltzmann transport formalism.[1] In the other extreme of models, thermally excited hops between high-energy localized vibrational modes are adopted to describe thermal diffusion in glassy materials, in addition to the low-frequency propagating modes. These localized modes do not propagate for long range but still carry heat and have a finite thermal diffusivity, which can be determined by the vibrational density of states (VDOS) and transition rates between them.[2] In the intermediate regime, the mixed nature of propagating and localized modes lends itself to the complexity where both the phonon propagating and hopping descriptions may both fail. Insights into the heat transport process in materials with intermediate level of disorder are very limited due to this complexity, and rely much on microscopic experimental characterization and atomistic simulations where the disorder could be controlled.

The roles of disorder and heat flux localization in defining thermal properties of materials are more significant in low-dimensional materials than in their bulk counterpart.[3, 4] The synthesis and characterization of two-dimensional (2D) materials have seen remarkable advances recently. Experimental evidences have demonstrated the existence of crystalline, amorphous regions and their interfaces in silica bilayers and graphene. Even their growth and shrinkage could be tuned and tracked under control.[5-9] These 2D materials thus provide an ideal platform to explore the effect of disorder in thermal conduction.[10, 11] To gain insights into the thermal transport in 2D materials with varying levels of disorder, we perform molecular dynamics (MD) simulations and analyze the results through theoretical models of heat transport in crystals and glasses. We find that as the level of disorder increases, the temperature dependence of thermal conductivity is turned over, from the signature behavior of crystal to that of the glass. The critical level of disorder is shown to correspond to the dominance of thermal diffusion from local mode hops over long-range phonon propagation, and the occurrence of heat flux localization.

**Results and Discussion**

2D silica films grow on metal substrates, forming both monolayer and 'bilayer' structures.[6]. In contrast to the monolayer films that are covalently bonded to the metal substrate, bilayers with two hexagonal silicon layers can be suspended on the substrate or isolated.[11] The bilayer structure are stable in both crystalline and vitreous phases with a corner-sharing $SiO_4$ network.[12] Recent studies not only identified the crystalline-vitreous interface of 2D silica bilayers, but also succeeded in exciting and imaging microscopic processes such as defect formation, structural deformation and phase transition in a controllable manner.[5, 10] Silicon pentagons and heptagons are the major types of defects in the vitreous phase, which is similar as its carbon monolayer analogy – graphene. Following these experimental evidences, we construct 2D hexagonal silica models with random 5/7/7/5 Stone-Wales defects that are distributed uniformed in the crystalline lattice, as illustrated in **Fig. 1**. We define the level of disorder $\alpha$ as the ratio between the number of non-hexagons and all the polygons or minimal rings.

The thermal conductivity $\kappa$ of a solid can be formulated at the continuum level through the kinetic relation $\kappa = cv_g l/d$, where $c$ is the specific heat, $v_g$, $l$ are the group velocity and mean free path of propagating phonons, and $d$ is the dimension of material. The value $v_g$ can be determined from the phonon dispersion or the elastic constant of the material. We calculate the mechanical response of 2D silica structures under tensile loads using the Tersoff interatomic potential.[13-15] The tensile stiffness of $Y$ as a function of the level of disorder $α$ is plotted in **Fig. 2(a)**, which can be fitted into a linear relation as $Y = 158.51 - 37.44α$ GPa. The thermal conductivity of 2D silica will thus be reduced as $α$ increases, due to the decrease in $v_g$, as well as reduction in $l$ that is expected to be significant due to the nonlocal effect of defects.[16] As a result, in contrast to the gentle dependence of $Y$ on $α$, the thermal conductivity is very sensitive to the presence of disorder. We calculate the in-plane thermal conductivity of the 2D silica $κ$ with different values of $α$, by using the Green-Kubo formula (see **Methods** for details). The temperature dependence of $κ$ is plotted in **Fig. 2(b)** for $T$ between 200 and 1000 K, where the quantum effect is not significant and thus classical MD simulations gives reliable predictions.[17] For hexagonal 2D silica crystal with $α = 0$, the value of $κ$ decreases with the temperature, in consistence with the general feature of heat conduction in crystals, which are evidenced in theoretical and experiment studies of graphene,[18] single- and poly-crystal silicon.[19,20] The major source of the thermal resistance here is the phonon-phonon interaction, which increases with the temperature.[21] As $α$ increases from 0 to 0.3, the $T$-dependence of $κ$ is significantly reduced because the translational crystalline symmetry is broken down and the phonon propagation is scattered by defects. As the level of disorder further increases beyond a critical value of $α_{cr} = 0.3$, the $T$-dependence is turned over – the value of $κ$ starts to increase with temperature and the $T$-dependence is enhanced as $α$ increases. This fact has been widely characterized in glassy solids such as amorphous silicon,[22] and bulk vitreous silica.[23]

The turnover of $T$-dependence in $κ$ indicates a transition of underlying heat transport mechanism. In order to elucidate the contribution of both propagating phonon and diffusive hops, we separate contributions from these two mechanisms with different nature in analyzing our MD simulation results.[24] Consider them as parallel processes we can write down the total thermal conductivity as

$$κ = κ_P + κ_D \qquad (1)$$

where $κ_P$ is the contribution from propagating phonon,[1] and can be evaluated as

$$\kappa_{\mathrm{P}} = \frac{1}{V}\int_0^{\omega_c} f(\omega)c(\omega)D(\omega)\mathrm{d}\omega \tag{2}$$

Here $V$ is the volume of material, $\omega$ is the frequency of vibrational modes. The integral is evaluated below the cut-off frequency $\omega_c$ for propagating modes. $f$, $c$, $D$ are the frequency-dependent, mode-specific VDOS, specific heat and thermal diffusivity. The value of $\omega_c$ is determined by the validity of Debye approximation in predicting the VDOS through $f(\omega) = 3V\omega^2/2\pi^2 v_s^2$. Here the phonon dispersion is simply assumed to be isotropic and linear and $v_s$ is the sound speed. From our MD simulation results, the VDOS is calculated from the Fourier transformation of the atomic velocity auto-correlation function (VACF). The results are close to the values we calculate from the elastic constants of the material as introduced earlier. Beyond $\omega_c$ the VDOS shows clear deviation from the $\omega^2$-dependence and higher-order terms appear (**Fig. S2**).

As has been done for amorphous silica and silicon,[22, 25-28] the diffusive contribution $\kappa_D$ in **Eq. (1)** can be calculated following Allen and Feldman's (AF) theory as[22, 29]

$$\kappa_{\mathrm{D}} = \frac{1}{V}\sum_{\omega_i > \omega_c} c(\omega_i)D(\omega_i) \tag{3}$$

Here the summation runs over diffusive mode $i$ specified by its frequency $\omega_i$. The AF formalism for the thermal diffusivity is[2]

$$D(\omega_i) = \frac{\pi V^2}{\hbar^2 \omega_i^2}\sum_{j\neq i}|S_{ij}|^2 \delta(\omega_i - \omega_j) \tag{4}$$

Here $S_{ij}$ is the heat current operator that measures the coupling between modes $i$ and $j$, and is determined by their frequencies and spatial overlap of mode vectors.[2, 22] $\hbar$ is the reduced Planck constant and $\delta$ is the Kronecker delta.

We now define a parameter $\eta = \kappa_D /(\kappa_P + \kappa_D)$ to measure the relative contribution of the diffusive heat transport, and calculate its value from AF theory. The results are summarized in **Fig. 3(a)**. We find that the value of $\eta$ increases from ~50% to ~90% with $\alpha$ increases up to 0.5 while the total thermal conductivity keeps decreasing. From the trend of $\eta$-$\alpha$ dependence, we can define a critical value $\alpha_{cr} = 0.3$ for the transition, which agrees well with the critical level of disorder characterizing the turnover of $T$-dependence in thermal conductivity. In **Fig. 3(b)**, we plot the thermal conductivities of 2D silica calculated from MD simulation results and normalized by $\kappa$ at $T = 300$ K, as well as the results predicted from the kinetic or AF formula with $\alpha = 0$ and 0.5. We find that the kinetic formula fits reasonably well the simulation results for hexagonal 2D silica crystals with $\alpha = 0$ while the AF model fails to predict well the MD simulation results for less disordered structure with $\alpha$ below 0.5.

In bulk materials where heat transfers in three-dimensional space, vibrational modes with frequency beyond a threshold value can be localized in a spatial span of a few lattice constants. While as the dimensionality is reduced to two or one, scaling theory predicts that all modes could be localized due to the presence of disorder, although this effect is much reduced compared to electrons due to the conduction through acoustic phonons and low-energy phonons.[3, 4, 30] The localization of phonon or vibrational modes can be characterized through the participation ratio $p$[31, 32]

$$p_\beta(\omega) = \frac{1}{3N_\beta} \frac{\left\{\sum_{i \in \beta}^{N_\beta} \sum_\mu [v_{i\mu}(\omega)]^2\right\}^2}{\sum_{i \in \beta}^{N_\beta} \sum_\mu [v_{i\mu}(\omega)]^4}, \quad \beta = \text{Si}, \text{O}, \mu = \{x, y, z\} \tag{5}$$

Here $\beta$ is the index for atom types, i.e. Si or O. $N$ is the total number of atoms. $\omega$ is the frequency of vibrational modes and $v_{i\mu}$ is the $i\mu$-component of the normalized eigenvector.[33] $p$ is of order unity when the mode is extended and becomes very small if the mode is spatially localized. Based on our MD simulations, we plot the $\omega$-dependence of $p$ for 2D silica with $\alpha = 0$ and 0.5 in **Fig. 4**. The results show that for $\alpha = 0$, the value of $p$ is large and most of the modes are extended. While for $\alpha = 0.5$, $p$ decreases significantly for modes with frequencies higher than 500 cm$^{-1}$ and most of the modes become localized. To quantitatively describe the localization of vibrational modes, we define a parameter $\gamma$ as the fraction of total number of modes that are localized. The criterion that a mode is localized if $p < 1/N^{1/2}$ is used following previous studies on the electron localization.[31, 34] The results for $\gamma$ are plotted in **Fig. 4** as a function of the level of disorder $\alpha$, which shows that $\gamma$ increases with $\alpha$ and exceeds 50% at $\alpha \sim 0.375$. This consistent value of $\alpha$ again validates our previous definition of critical level of disorder.

To gain more insights into the localization of thermal energy transfer in disorder 2D materials, we calculate the atomistic heat flux from our nonequilibrium molecular dynamics (NEMD) simulations of 2D silica, which is defined from the expression $\mathbf{J}_i = -\mathbf{S}_i \mathbf{v}_i$, where $\mathbf{S}_i$ is the atomic stress tensor and $\mathbf{v}_i$ is the velocity vector of atom $i$. The amplitude of local flux $J$ is averaged over 5 ps during the steady state in simulations. We plot the normalized spatial distribution of heat flux $(J_i - \langle J \rangle)/J_{max}$ in **Fig. 5(b)**. The results show that for $\alpha = 0$, the spatial distribution is relatively uniform, while as $\alpha$ increases to 0.1, the heat flux starts to be perturbed by the presence of Stone-Wales defects. However, the perturbation is still weak and localized. For 2D silica with $\alpha > 0.3$, our simulation results demonstrate strong scattering of the heat flux and the pattern of $J$ becomes localized spatially. To quantify the level of spatial localization, we define a roughness of heat flux distribution as

$$R_{RMS} = \sqrt{\sum_{i=1,N} \mathbf{J}_i^2 / N} \tag{6}$$

and a localization factor (LF) following the approach in Ref.[35]. The results plotted in **Fig. 5(c)** again show an almost bilinear dependence on $\alpha$ with a turning point at $\alpha = 0.3$, which aligns with our previous discussion on the mechanistic transition.

## Conclusion

Our MD simulation results show that thermal transport in 2D silica with Stone-Wales type of disorder becomes dominated by the thermal hopping between localized modes with $α > α_{cr} = 0.3$, which explains a turnover observed in the temperature dependence of the thermal conductivity. The thermal conductivity decreases with temperature at $α < α_{cr}$, and increases at $α > α_{cr}$, and the *T*-dependence is weakened as $α$ approaches $α_{cr}$ for both cases. The determination of the critical level of disorder is validated from both the analysis of *T*-dependence turnover, dominance of diffusive contribution of thermal conductivity, the participation ratio of localized modes and the spatial localization of heat flux, which all can be used as a good indicator for the transition of heat transport mechanism. The localization of heat flux at high level of disorder results in dramatic reduction in the thermal conductivity[36]. This conclusion is expected to be general for other 2D materials, which is validated by our extended studies on graphene.[16, 37] However in our MD simulations, the non-hexagons in graphene constructed by the Stone-Wales rotation are not stable at elevated temperature over 500 K, and thus the exploration of the *T*-dependent thermal conduction is limited.

There are a few theoretical models that work well for the limiting cases of crystal and glass, by introducing the mechanisms of propagating phonons or thermal hopping between localized modes. However, the applicability of these two classes of models need to be justified for materials with intermediate level of disorder, where either the polarization and group velocity of localized modes or the hopping and diffusivity of extended propagating modes cannot be well defined[17] but are capable to carry heat flux. The recently isolated 2D materials include a wide spectrum of crystalline lattices and types of disorder or defects, and thus provide an ideal test-bed to explore the detailed microscale dynamics related to the thermal energy transfer.

## Methods

We calculate the thermal conductivity $κ$ of two-dimensional (2D) silica from equilibrium molecular dynamics (EMD) simulations, using the linear response theory based Green-Kubo formulism,[38, 39] which applies for systems in thermal equilibrium where heat flux fluctuates around zero. $\kappa$ could thus be expressed as an integration of the heat flux operator multiplied by a prefactor

$$\kappa_{xy} = \frac{1}{Vk_{\mathrm{B}}T^2} \int_0^{\tau_c} \langle \mathbf{J}_x(\tau) \cdot \mathbf{J}_y(0) \rangle \mathrm{d}\tau,$$

where $T$ is the temperature of system, $k_{\mathrm{B}}$ is the Boltzmann constant, and $V$ is the system volume that was defined here as the area of 2D silica multiplied by its nominal thickness 6.3 Å. The upper limit of time integration $\tau_c$ needs to be long enough so the current-current correlation function decays to zero.[40] $J_x$ and $J_y$ are the heat current operators in the $x$ and $y$ directions, and the angular bracket represents the ensemble average,[41] namely the heat flux autocorrelation function (HFACF). The heat flux $\mathbf{J}$ of the system was computed from the expression $\mathbf{J} = (\Sigma_i e_i \mathbf{v}_i - \Sigma_i \mathbf{S}_i \mathbf{v}_i)/V$, where $e_i$, $\mathbf{v}_i$ and $\mathbf{S}_i$ are the total energy, velocity vector, and stress tensor of each atom $i$, respectively. We first integrate HFACF with an integration time $\tau$, then obtain the relation between $\kappa$ and $\tau$. The decorrelation time for the heat flux is typically on the scale 10 ps for our models, and thus converged results for $\kappa$ could be extracted when $\tau > \tau_c$ in the simulations. We evaluate $\kappa$ as the mean value of $\kappa_{xx}$ and $\kappa_{yy}$ that may differ in the finite system under simulation though, where the maximum difference between $\kappa_{xx}$ and $\kappa_{yy}$ in our simulations is less than $0.1\kappa$.

All simulations are performed using the large-scale atomic/molecular massively parallel simulator (LAMMPS).[42] Periodic boundary conditions are applied to a 2D supercell of silica, and in order to minimize the size effect, we use almost the same supercell size in the MD simulations (23.3×22.7 nm$^2$). The Tersoff potential is used for the interatomic interactions between silicon and oxygen atoms, which has been demonstrated to provide an excellent prediction for the phonon dispersion of silica compared to experimental measurements.[13-15] In our MD simulations, atomic structures of 2D silica bilayers are equilibrated at $T = 300$ K, where the quantum correction is negligible,[43] by coupling to a Nosé-Hoover thermostat for 200 ps. The time step is 0.1 fs. The structures are further equilibrated in a microcanonical ensemble for 50 ps before the thermal conductivity is calculated from the EMD simulations in the same ensemble. The atomic positions and velocities are collected along the simulations to evaluate the heat flux and its autocorrelation functions. The thermal conductivity is finally calculated through the Green-Kubo formula and averaged over eight independently sampled runs for each structure. This approach is validated by comparing our MD simulation results with recent experimental measurements (**Supplementary Fig. 1**). To obtain the spatial distribution of heat flux in the sample, we also carry out non-equilibrium molecular dynamics (NEMD) following the idea of Müller-Plathe, where energy exchange is driven between the heat source and sink to create a steady heat flux.[44]

## Acknowledgments

This work was supported by the National Natural Science Foundation of China through Grant 11222217 and 11472150, the State Key Laboratory of Mechanics and Control of Mechanical Structures (Nanjing University of Aeronautics and Astronautics) through Grant No. MCMS-0414G01. The computation was performed on the Explorer 100 cluster system at Tsinghua National Laboratory for Information Science and Technology.

**Figures and captions**

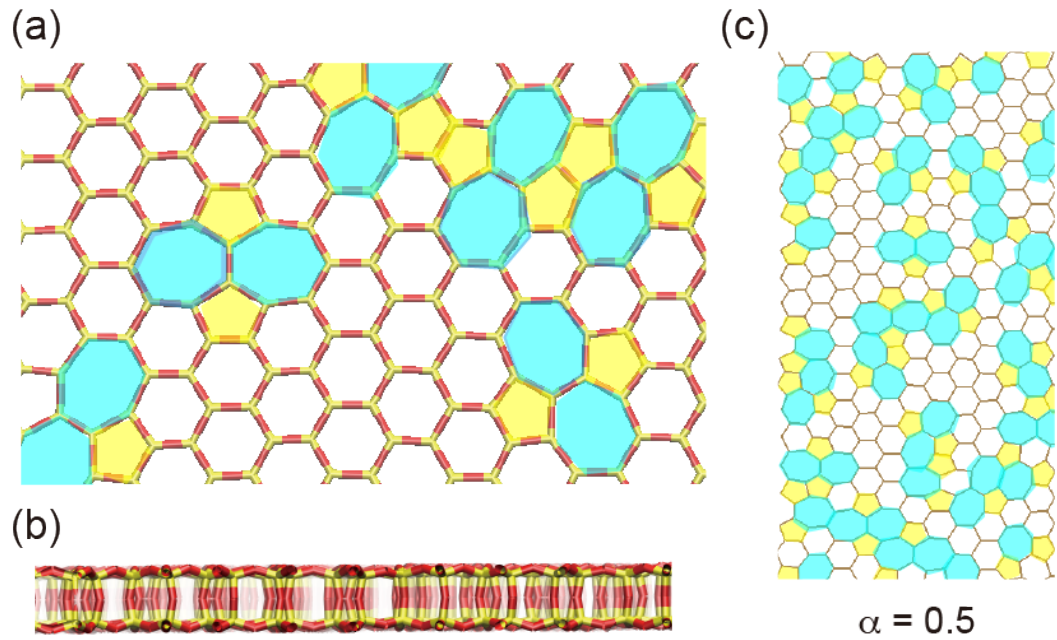

**Figure 1.** Models of two-dimensional silicon dioxides bilayer. (a, b) Top and side views of 2D silica topological defects (pentagons, heptagons) filled with yellow/light bule colors to distinguish from hexagons. (c) Atomic structures of 2D silica with the level of disorder $\alpha = 0.5$, which is defined by the ratio between non-hexagons and all polygons in the lattice.

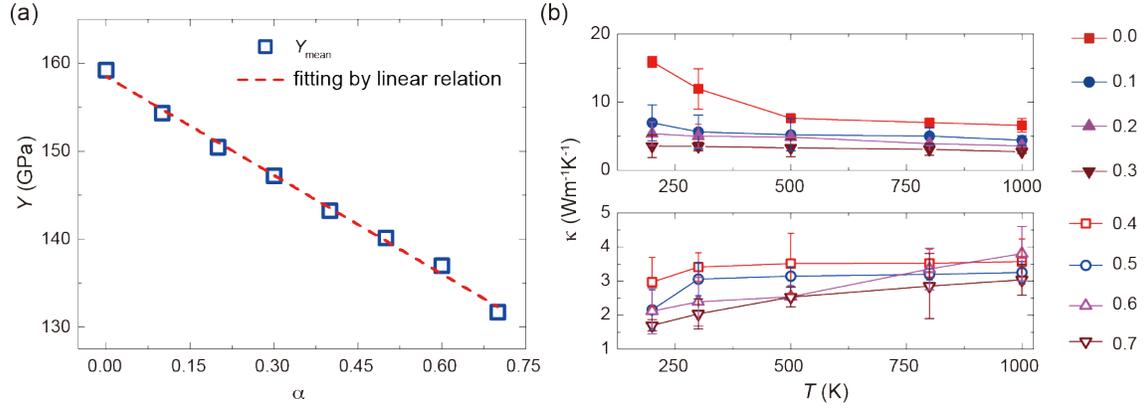

**Figure 2.** (a) Tensile stiffness of 2D silica plotted as a function of the level of disorder $\alpha$, fitted using a linear function $Y = 158.51 - 37.44\alpha$ GPa. (b) Temperature dependence of the thermal conductivity $\kappa$ of 2D silica, plotted as a function of $\alpha$. There exist a critical level of disorder $\alpha_{cr} = 0.3$. $\kappa$ decreases with $T$ for $\alpha \leq \alpha_{cr}$, and increases with $T$ for $\alpha > \alpha_{cr}$.

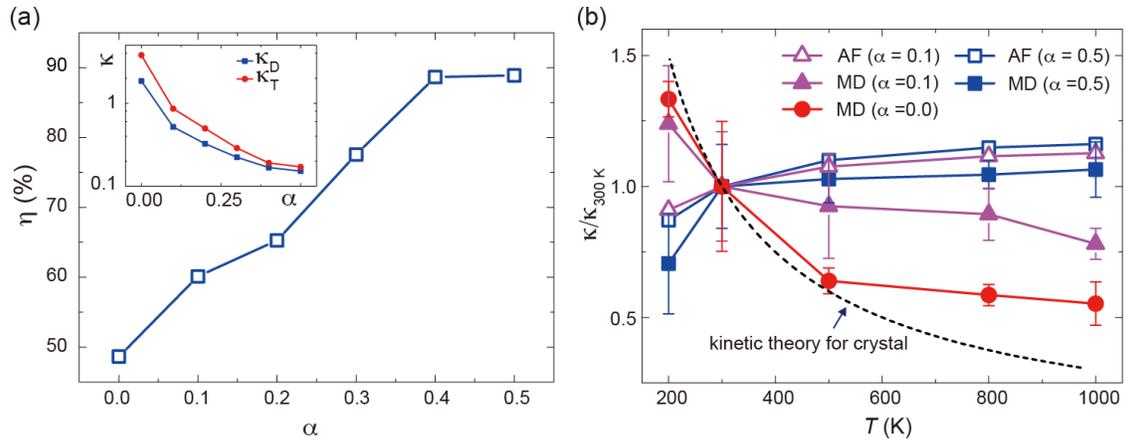

**Figure 3.** (a) The contribution of thermal diffusion in the total conductivity $\eta$, which is resulted from hops between localized vibrational modes. The inset plots both total thermal conductivity and the diffusive part as calculated from the Allen-Feldman theory. (b) Comparison between theoretical models and MD simulation results for crystalline and amorphous 2D silica with $\alpha = 0.1$ and $0.5$.

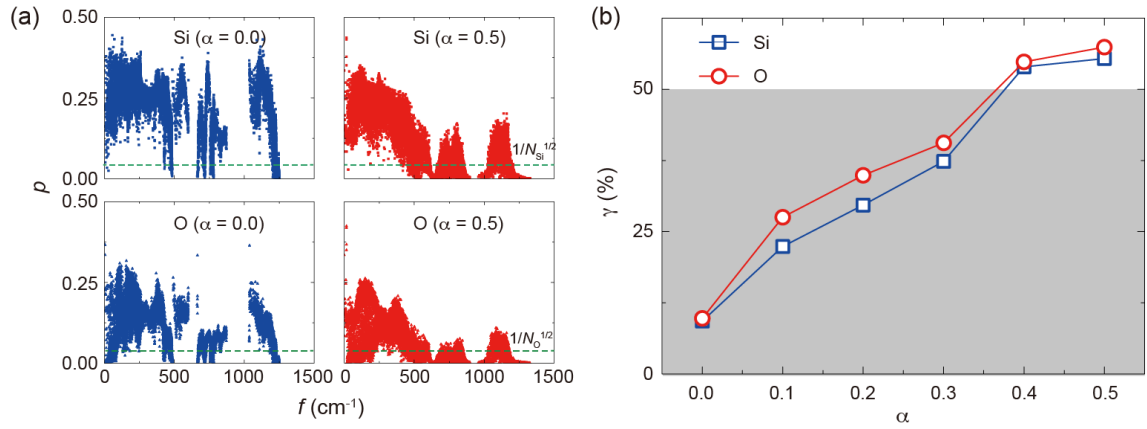

**Figure 4.** (a) The participation ratio $p_{Si}$ and $p_O$ for 2D silica with $\alpha = 0.0$ and 0.5. The dash lines indicate the criterion of vibrational mode localization ($p < 1/N^{1/2}$). (b) The fraction of localized vibrational modes plotted as a function of $\alpha$.

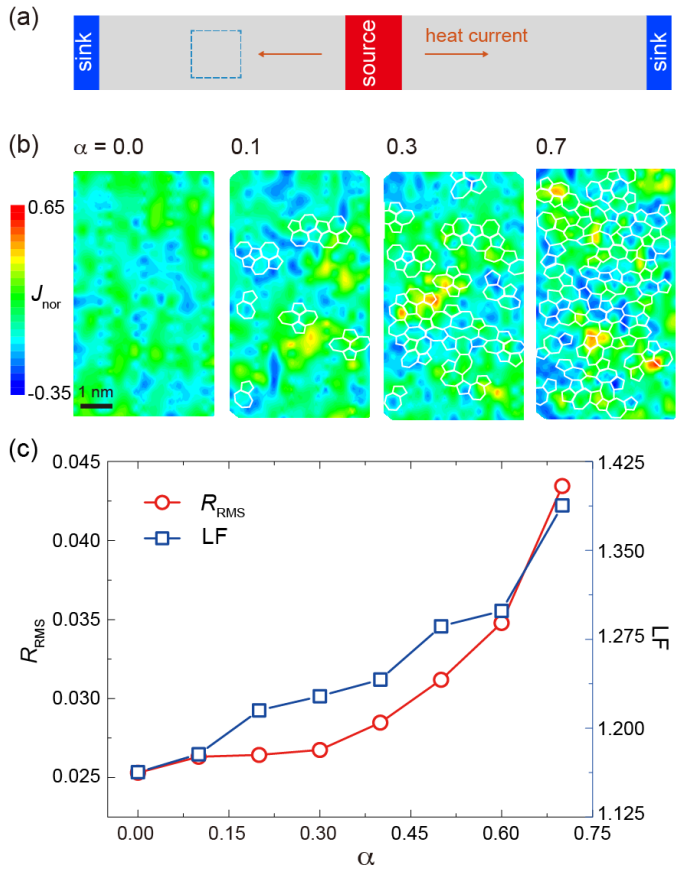

**Figure 5.** (a) Illustration of thermal transport simulation setup for 2D silica. Here we select a window to analyze the spatial distribution of heat flux in the steady state. (b) Spatial distribution of the amplitude of heat flux $J$ (in $Wm^{-2}$) for 2D silica with $\alpha$ = 0.0, 0.1, 0.3, and 0.7. The scale bar is 1 nm. (c) The spatially averaged $J$ plotted as a function of $\alpha$.